\documentclass[sigconf]{acmart}

\AtBeginDocument{%
  }

\copyrightyear{2025}
\acmYear{2025}
\setcopyright{acmlicensed}
\acmConference[WWW Companion '25]{Companion Proceedings of the ACM Web Conference 2025}{April 28-May 2, 2025}{Sydney, NSW, Australia}
\acmBooktitle{Companion Proceedings of the ACM Web Conference 2025 (WWW Companion '25), April 28-May 2, 2025, Sydney, NSW, Australia}
\acmDOI{10.1145/3701716.3715475}
\acmISBN{979-8-4007-1331-6/2025/04}

\usepackage{multirow}
\usepackage{array}
\usepackage{float}
\usepackage{stfloats}
\usepackage{microtype}
\usepackage{graphicx}
\usepackage{subfigure}
\usepackage{booktabs} 
\usepackage{algorithm}
\usepackage{algpseudocode}
\usepackage{amsmath}
\usepackage{amsfonts}
\usepackage{colortbl}
\usepackage{arydshln} 
\begin{document}

\title{Conf-GNNRec: Quantifying and Calibrating the Prediction Confidence for GNN-based Recommendation Methods}



\author{Meng Yan}
\affiliation{%
  \institution{Xidian University}
  \city{Xi'an}
  \country{China}}
\orcid{0000-0001-8478-4823}
\email{mengyan@stu.xidian.edu.cn}

\author{Cai Xu}
\affiliation{%
  \institution{Xidian University}
  \city{Xi'an}
  \country{China}}
\orcid{0000-0002-7191-7348}
\email{cxu@xidian.edu.cn}

\author{Xujing Wang}
\affiliation{%
 \institution{Xidian University}
 \city{Xi'an}
 \country{China}}
\orcid{0009-0003-5336-8845}
\email{xjwong@stu.xidian.edu.cn}

\author{Ziyu Guan}
\authornote{Corresponding author}
\affiliation{%
  \institution{Xidian University}
  \city{Xi'an}
  \country{China}}
\orcid{0000-0003-2413-4698}
\email{zyguan@xidian.edu.cn}

\author{Wei Zhao}
\affiliation{%
  \institution{Xidian University}
  \city{Xi'an}
  \country{China}}
\orcid{0000-0002-9767-1323}
\email{ywzhao@mail.xidian.edu.cn}

\author{Yuhang Zhou}
\affiliation{%
 \institution{Communication University Of China}
  \city{Beijing}
  \country{China}}
\orcid{0009-0000-4460-2001}
\email{yhz@mails.cuc.edu.cn}

\renewcommand{\shortauthors}{Meng Yan et al.}
\renewcommand{\shorttitle}{Conf-GNNRec}

\begin{abstract}
Recommender systems based on graph neural networks perform well in tasks such as rating and ranking. However, in real-world recommendation scenarios, noise such as user misuse and malicious advertisement gradually accumulates through the message propagation mechanism. Even if existing studies mitigate their effects by reducing the noise propagation weights, the severe sparsity of the recommender system still leads to the low-weighted noisy neighbors being mistaken as meaningful information, and the prediction result obtained based on the polluted nodes is not entirely trustworthy. Therefore, it is crucial to measure the confidence of the prediction results in this highly noisy framework. Furthermore, our evaluation of the existing representative GNN-based recommendation shows that it suffers from overconfidence. Based on the above considerations, we propose a new method to quantify and calibrate the prediction confidence of GNN-based recommendations (Conf-GNNRec). Specifically, we propose a rating calibration method that dynamically adjusts excessive ratings to mitigate overconfidence based on user personalization. We also design a confidence loss function to reduce the overconfidence of negative samples and effectively improve recommendation performance. Experiments on public datasets demonstrate the validity of Conf-GNNRec in prediction confidence and recommendation performance.
\end{abstract}






\maketitle

\begin{figure}[H]
\centering
\includegraphics[width=0.46\textwidth]{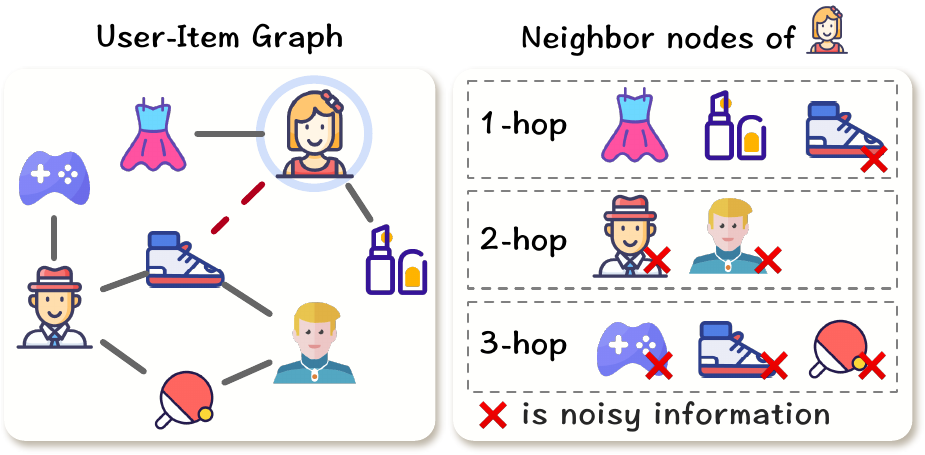} 
\caption{Illustration of the propagation of noise in the GNN-based recommendation. Sparse interactions result in 2-hop and 3-hop neighbor nodes being noisy information.}
\label{fig1}
\end{figure}

\section{Introduction}
Recommender systems play an important role in prediction making of various social media such as e-commerce \cite{yan}, entertainment \cite{4,15}, and social networks\cite{5,lu3}. 
Interactions between users and items can be naturally constructed as graphs, and graph structures can represent the higher-order connections that users establish with other users via common items.
Graph Neural Network (GNN), by virtue of its ability to model interactions and capture higher-order dependencies, has become the powerful framework for recommender systems\cite{7}.

Most of the GNN-based recommendation methods make improvements in all or some of the following three steps: 1) \textbf{Graph Construction. }Transform the user-item interactions in the recommender system into a graph structure, which can introduce side information to add additional nodes and edges. 2) \textbf{Message Propagation. } Propagate and aggregate users and items information on the graph based on GNN. 3) \textbf{Recommendation.} Calculate interaction probabilities or ratings based on the final embeddings of users and items.
In practical recommendation scenarios, noise such as user misuse and malicious advertisements will gradually accumulate through the message propagation mechanism. 
Currently, there have been research attempts to reduce the weight of noise through attention \cite{6,11,12}, confidence\cite{2,13,14}, and other methods in the propagation process. 
Since node features are mainly complemented by the neighbor nodes, severe sparsity in recommendation can lead to low weighted neighbors also being recognized as meaningful information, which can lead to layer-by-layer aggregation of noise and pollution of node features\cite{3,lu2}. 
As shown in Figure \ref{fig1}, a female user accidentally clicked on a pair of shoes, and the shoe preference belongs to noise information. Meanwhile, other items that the user interacts with are very unpopular, the noise of this pair of shoes will be more aggregated during the message propagation process, leading to further propagation of its noise impact.

Recommendation results usually rely on similarity calculations between user nodes and item nodes, and the interference of noise can lead to predictions that are not fully trustworthy. Therefore, it is crucial to measure the confidence of predictions in this high noise framework. In other words, we need to assess the trustworthiness of the result through the probability, i.e., confidence level, associated with the prediction result. In general, the predicted probability at the output layer represents the confidence for the prediction. 
Therefore, it naturally raises the fundamental question: \textit{does the confidence given directly by a GNN-based recommendation model match the accuracy of the prediction?}

To explore this question, we evaluate the relationship between confidence and accuracy for lightGCN \cite{7} and KGCL \cite{8} on publicly available datasets respectively, where confidence is the activation function output of the last layer of the model. Surprisingly, confidence in existing GNN-based recommendations is significantly flawed, with the confidence being much higher than the accuracy. These probabilities obtained from the activation layer only take into account the relative magnitude of the outputs and do not truly reflect the level of confidence the model has in the predicted results. More importantly, the overconfidence confidence that the model assigns to the prediction results can harm the user experience. 
In high-risk scenarios such as finance and healthcare, wrong recommendations can cause significant loss of life and property.

Based on the above challenges, we propose a new method to quantify and calibrate the prediction confidence of GNN-based recommendations (Conf-GNNRec), which adopts a post-calibration strategy without other additional inputs. Specifically, we design a rating calibration method that dynamically adjusts excessive ratings to mitigate overconfidence based on user personalization. In addition, we design a new confidence loss function that effectively reduces the high confidence of negative samples, further bridging the gap between recommendation accuracy and confidence.
The main contributions of this work can be summarized as follows: 

\begin{itemize}
\item To the best of our knowledge, we are the first to focus on prediction confidence in GNN-based recommendation and to point out that existing methods are overconfident in predicting results.
\item We propose a new prediction confidence measure method that can improve recommendation performance while mitigating overconfidence.
\item We compare the performance changes of representative GNN-based recommendation methods before and after applying Conf-GNNRec on a public dataset, and the experimental results demonstrate the effectiveness of Conf-GNNRec.
\end{itemize}

\section{Method}
\subsection{Problem Statement}

In recommender systems, user-item interactions can be naturally constructed as graphs $\mathcal{G=\{ V, E }\}$. The $\mathcal{V} = \mathcal{V}_U \cup \mathcal{V}_I$ is the node set with $(N+M)$ nodes,  where $\mathcal{V}_U=\{v_1,v_2,...,v_N\}$ represents the user set, and $\mathcal{V}_I=\{v_{N+1},v_{N+2},...,v_{N+M}\}$ represents the item set. $\mathcal{E} = \{(v_ u, v_i)| y_{ui}=1 \}$ is the edge set, where $y_{ui} = 1$ indicating that there is a implicit interaction between $v_u \in \mathcal{V}_U$ and $v_i \in \mathcal{V}_I$. 
Given a GNN-based recommendation model $f_{\theta}$ where $\theta$ is the learnable parameters. 

\subsection{GNN-based Recommendation Confidence}
The GNN-based recommendation model firstly defines the initialised embedding features of the nodes: $ \mathbf{h}_u^{(0)} = Embed(v_u)$ and $ \mathbf{h}_i^{(0)} = Emb(v_i)$, where the $Emb(\cdot)$ is the embedding function. Then, it aggregates the neighbor information of users and items and updates the representation. After the $L$-th layer of the network, it can be represented as:

\begin{equation}
    \mathbf{h}_*^{(L+1)} = \text{Update}^{(L)}(\mathbf{h}_*^{(L)},
    \text{Aggregator}^{(L)}(\{ \mathbf{h}_{v_i}^{(L)}, 
    \forall v_i \in \mathcal{N}(v_*) \}))
\end{equation}
where $\text{Update}^{(L)}$ and $\text{Aggregator}^{(L)}$ represent the function of aggregation and update at the $L$-th layer, respectively.$ \mathcal{N}(v_*)$ indicates the neighbor nodes of $v_*$. We consider it as the final representation of users and items to predict interactions or ratings:
\begin{equation}
    r_{u,i} = score(\mathbf{h}_u^{(L+1)}, \mathbf{h}_i^{(L+1)})
\end{equation}

\begin{figure}[t]
\centering
\includegraphics[width=0.45\textwidth]{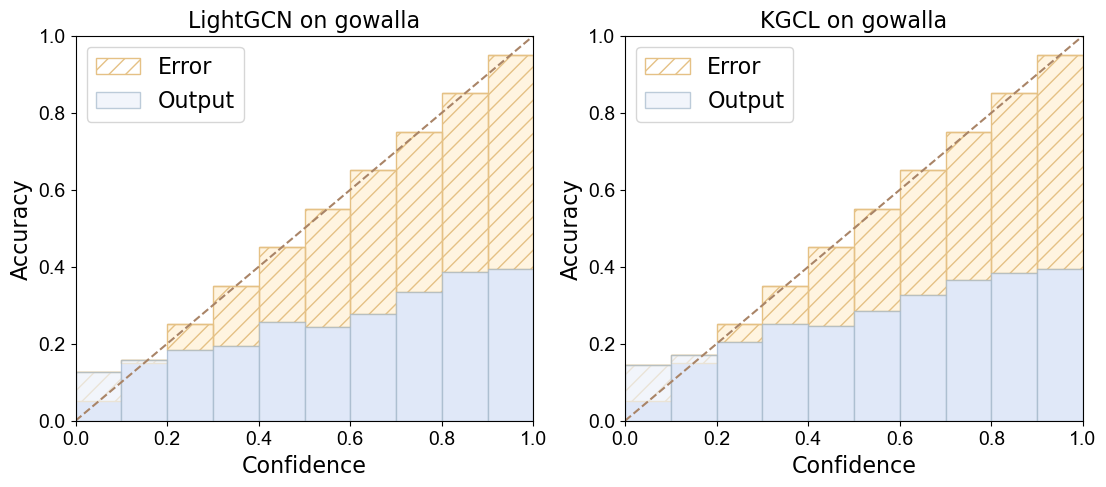} 
 \caption{Reliability diagrams for LightGCN (left) and KGCL (right). The blue bar is the output of model, the yellow bar is the output in the perfect case. Any deviation from a perfectly diagonal (i.e., yellow slashes) represents the miscalibration.}
\label{fig2}
\end{figure}

To facilitate comparison of results, the results are fed into the normalization layer: $z_u = [\sigma(r_{u,1}),\sigma(r_{u,2}),...,\sigma(r_{u,M})]$, where $\sigma()$ is the normalization function. 
For the user $u$, $\hat{y}_{u,i} = \arg \max_{i} z_{u,i}$ indicates the prediction and $\hat{p}_{u,i} = \max_i z_{u,i}$ indicates the corresponding confidence. 
We believe that perfect confidence should satisfy the following equation:
\begin{equation}
    \mathbb{P}(\hat{y}_{u,i} = y_{u,i} | \hat{p}_{u,i} = {p}) = p, 
    \forall p \in [0,1]
\end{equation}
For the top-k task, we use the mean value instead. Next, we visualize the relationship between confidence and accuracy using two representative GNN-based recommendations (lightGCN and KGCL) as an example, as shown in figure \ref{fig2}. Specifically, we divide the confidence interval $[0,1]$ into ten equal bins with reference to \cite{17}, group the results into the corresponding bins based on the confidence, and then compute the average accuracy of each bin. For a perfect model, each bin's accuracy is exactly equal to the confidence, i.e., all blue bars match the brown diagonal. It is clear that the model outputs a confidence level that is too high and that the confidence level does not match the accuracy (yellow slashes).

\subsection{Conf-GNNRec}
In this section, we propose the rating calibration method to reset the ratings. The function should 1) adapt to different user interaction habits. 2) be non-linear, and 3) maintain recommendation performance. 
Based on the above objectives, we design a segmented processing strategy to achieve rating calibration. A low rating usually indicates that the user has little interest in the item, and the original values can be retained directly. High ratings represent strong user preferences but may also overstate the weights of items.
Therefore, we use logarithmic transformation to nonlinearly compress the portion that exceeds the mean value while retaining the relative ranking information. We reset the ratings $r'_u$ as follows:

\begin{equation}
    r'_u = [rc(r_{u,1}),rc(r_{u,2}),...,rc(r_{u,M})]
\end{equation}

\begin{equation}
    rc(r_{u,i}) = r_{u,i} \cdot \mathbb{I}(r_{u,i} \le \bar{r}_{u}) + 
        (\tau \log ({r_{u,i}-\bar{r}_{u}})+ \bar{r}_{u})
        \cdot \mathbb{I}(r_{u,i}>\bar{r}_{u})
\end{equation}
where $rc()$ is responsible for calibrating overconfidence. $\mathbb{I}()$ is the indicator function that outputs 1 when the condition is true, and $\bar{z}_{u}$ is the mean of all elements in $z_u$. $\tau$ is the temperature factor controlling the smoothness of ratings. 
So we can get a new probability distribution: $p'_u = [\sigma(r'_{u,1}),\sigma(r'_{u,2}),...,\sigma(r'_{u,M})]$. 

In addition, we separately penalize the items that are wrongly predicted but with high confidence, which not only effectively mitigates the over-confidence but also improves the accuracy of recommendation to some extent. It can be flexibly applied to all GNN-based recommender systems. The proposed regularization loss is as follows:
\begin{equation}
    \mathcal{L}_{rcREC} = \sum_{u=1}^{N} \sum_{i=1}^{M}(1-p_{u,i}) 
    \hat{p}_{u,i} log(1+\exp(\hat{p}_{u,i}))
\end{equation}
where $\hat{p}_{u,i} log(1+\exp(\hat{p}_{u,i})$ increases the penalty for high confidence negative samples and increases nonlinearly with confidence. It also mitigates the effect of low confidence errors so that the model will make predictions about items more cautiously.
\section{Experiment}
In this section, 
we present the datasets, baseline methods, and evaluation protocols.
Subsequently, we apply Conf-GNNRec on different baseline models and compare the performance before and after, the related code\footnote{https://github.com/FairyMeng/Conf-GNNRec} is publicly available. Finally, we visualize the effect of the improvement in the confidence.
\subsection{Settings}

\subsubsection{\textbf{Datasets}}
We perform experiments on three public datasets collected from different real-life platforms: Gowalla for location recommendation, Yelp2018 for restaurant recommendation, and Amazon-Book for product recommendation. Table \ref{data}  presents the statistical information of our experimented datasets.
\begin{table}[ht]
    
    \caption{Statistics of Datasets}
    \label{data}
    \centering
    \begin{tabular}{lcccc}
    \toprule  
        \textbf{Datasets} &\textbf{User} &
        \textbf{Item} & \textbf{Interaction} & \textbf{Density} \\
        
        \midrule
        
        Gowalla     & 29,558 & 40,981 & 1,027,370 & 0.00084 \\
        Yelp2018    & 31,668 & 38,048 & 1,561,406 & 0.00130 \\
        Amazon-Book & 52,643 & 91,599 & 2,984,108 & 0.00062 \\
       
        \bottomrule
    \end{tabular}

\end{table}

\subsubsection{\textbf{Baseline}}

We apply Conf-GNNRec to the following baselines and compare them:
\begin{itemize}
\item  LightGCN\cite{7}. This is a GCN-based recommendation method that simplifies the convolution operations during message passing among users and items.
\item KGAT\cite{9}. This model designs an attentive message passing scheme over the knowledge-aware collaborative graph for embedding fusion. Setting weights for different neighbor nodes based on the attention mechanism.
\item MVIN\cite{10}. It is a multi-view item embedding method based on GNN. Information from both the user and the entity side is considered to learn feature embeddings of items.
\item KGCL\cite{8}. This is a general framework that introduces knowledge graphs to guide recommender systems in denoising information as it is aggregated.
\end{itemize}

\subsubsection{\textbf{Evaluation Protocols}}
For all of the above methods, we use open source code and uniform datasets. We randomly split the dataset in the ratio of 7:1:2 to build the training, validation and test sets. Moreover, it is ensured that each user has at least one item in the test set. We followed the authors' advice on the method for the hyperparameters used in the baseline model and reported the results under the optimal hyperparameter settings. 
In particular, KGCL additionally introduces a knowledge graph, resulting in a recommended dataset inconsistent with Table \ref{data}, and we conduct the experiment according to the authors' settings. 
For the performance evaluation, we use two representative metrics: $Precision@N$ and $Accuracy@N$ are used to evaluate the accuracy of top-N recommended items. Average evaluation results across all users in the test set are reported with $N = 20$ by default.

\begin{table}[t]
    
    \centering
    \renewcommand{\arraystretch}{1.2}
    \caption{Overall performance on Recommendation\label{tab:results_limited}}
	\resizebox{1\linewidth}{!}
{
    \begin{tabular}{l  cc  cc  cc}
    \hline
    \toprule
    \textbf{\textsc{Method}} & \multicolumn{2}{c}{\textbf{Gowalla}} & \multicolumn{2}{c}{\textbf{Yelp2018}} & \multicolumn{2}{c}{\textbf{Amazon-Book}}\\
     ~+\textbf{\textsc{Conf-GNNRec}} &  \textsc{Pre.} & \textsc{Acc.} & \textsc{Pre.} & \textsc{Acc.} &  \textsc{Pre.} &  \textsc{Acc.}\\

    \midrule

	LightGCN
	& 5.267 & 15.327 & 2.716 & 4.313 & 3.295 & 7.305  \\
	\hdashline[4pt/4pt]
	\rowcolor[HTML]{edf2fa}~+Conf-GNNRec & \textbf{5.386} & \textbf{15.597}  & \textbf{2.864}  & \textbf{4.496} & \textbf{3.356} & \textbf{7.418} \\
	\rowcolor[HTML]{fcedf2}~~\emph{Improve} (\%) 
	& \textit{2.259} & \textit{1.762} & \textit{5.449} & \textit{4.243} & \textit{1.851} & \textit{1.547} \\

	\midrule
	
	KGAT
	& 5.452 & 15.838 & 2.743  & 4.374 & 3.273 & 7.393  \\ 
	\hdashline[4pt/4pt]
	\rowcolor[HTML]{edf2fa}~+Conf-GNNRec & \textbf{5.577} & \textbf{16.225} & \textbf{2.882} & \textbf{4.512} & \textbf{3.346} & \textbf{7.547} \\
	\rowcolor[HTML]{fcedf2}~~\emph{Improve} (\%) 
	& \textit{2.293} & \textit{2.443} & \textit{5.067} & \textit{3.155} & \textit{2.230} & \textit{2.083} \\
	
	\midrule
	MVIN
	& 5.527 & 15.935 & 2.883 & 4.419 & 3.401 & 7.428  \\ 
	
	\hdashline[4pt/4pt]
	\rowcolor[HTML]{edf2fa}~+Conf-GNNRec & \textbf{5.652} & \textbf{16.448} & \textbf{3.011} & \textbf{4.616} & \textbf{3.489} & \textbf{7.604} \\
	\rowcolor[HTML]{fcedf2}~~\emph{Improve} (\%) 
	& \textit{2.262} & \textit{3.219} & \textit{4.093} & \textit{4.458} & \textit{2.587} & \textit{2.369} \\

	\midrule

        KGCL
        & 5.524  & 16.362 & 2.758 & 4.832 & 3.618 & 7.829 \\
        \hdashline[4pt/4pt]
        \rowcolor[HTML]{edf2fa}~+Conf-GNNRec & \textbf{5.644} & \textbf{16.871} & \textbf{2.942} & \textbf{5.093} & \textbf{3.735} & \textbf{8.139}\\
        \rowcolor[HTML]{fcedf2}~~\emph{Improve} (\%)
        & \textit{2.172} & \textit{3.111} & \textit{6.672} & \textit{5.401} & \textit{3.234} & \textit{3.960}  \\
	
	
    \bottomrule
    \hline
    \end{tabular}
    \label{rec-out}
}
\end{table}

\subsection{Overall Performance}
\subsubsection{\textbf{Recommendation}}
We compare the performance changes of the four recommendation models before and after the application of Conf-GNNRec and show the magnitude of improvement brought by Conf-GNNRec. 
Table \ref{rec-out} shows the experimental results. It can be clearly observed that the application of Conf-GNNRec improves all models and datasets. 
In particular, the improvement of Conf-GNNRec is more significant in complex models where more information is introduced (e.g., KGCL) and datasets with high density (e.g., Yelp2018). 
The above illustrates the effectiveness of our model design in reducing false predictions generated by overconfidence.

\subsubsection{\textbf{Confidence}}
\begin{figure}
\centering
\includegraphics[width=0.45\textwidth]{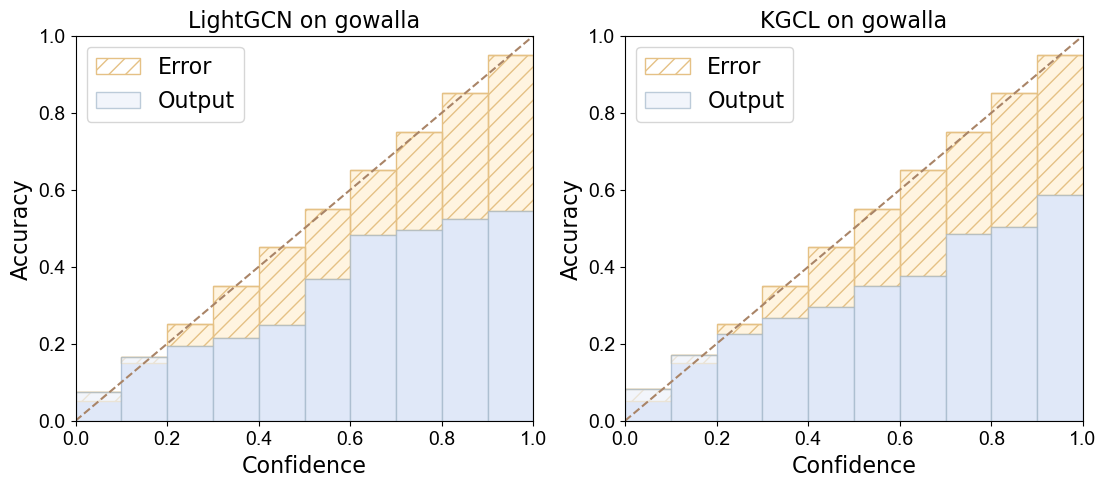} 
\caption{Reliability diagrams for LightGCN (left) and KGCL (right) after applying Conf-GNNRec.}
\label{fig3}
\end{figure}
In this section, we visualize the relationship between the prediction confidence and the accuracy in lightGCN and KGCL on Gowalla, as shown in figure \ref{fig3}. All settings are consistent with those in 3.1.2. By comparing figure \ref{fig2} and figure \ref{fig3}, it can be clearly seen that the model significantly reduces the error between confidence and accuracy. The KGCL model exhibits smoother relational consistency. In addition, as the confidence level increases, the accuracy of the recommender system shows a significant increase, which is in accordance with our intuitive understanding: the prediction with higher confidence tends to correspond to a more reliable recommendation result. 
However, we also recognize that achieving fully accurate confidence predictions in recommender systems is nearly impossible. This is mainly attributed to the high degree of personalization, uncertainty in user preferences, and the inevitable inclusion of noisy interactions (e.g., user incidental clicking behavior). Therefore, the goal of a trusted recommender system should focus on reducing the error between the prediction confidence and the actual accuracy, thus improving the credibility of the model and the user experience.

\section{Conclusion}
In this paper, we explore the relationship between confidence and accuracy of GNN-based recommender systems. By analysing existing approaches, we find overconfidence in the models. To address this issue, we propose a new prediction confidence measure and design a loss function that significantly mitigates overconfidence while improving recommendation performance. We compare the performance changes of representative GNN-based recommendation methods before and after applying Conf-GNNRec on a public dataset, and the experimental results demonstrate the effectiveness of Conf-GNNRec.

It is worth noting that the research on prediction confidence for GNN-based recommendation is in its infancy. We plan to use Bayesian to improve the trustworthiness of recommendation results further. Meanwhile, we will endeavor to extend to other recommendation frameworks (e.g., large language models) for exploring a wider range of application scenarios and potential value.

\begin{acks}
This research was supported by the National Natural Science Foundation of China under Grants 62133012, 62472340, 62425605, 62303366. Key Research and Development Program of Shaanxi under Grant 2024CY2-GJHX-15.
\end{acks}

\bibliographystyle{ACM-Reference-Format}
\bibliography{sample}


\end{document}